\documentclass[a4paper,fleqn,usenatbib]{mnras}

\usepackage[T1]{fontenc}
\usepackage{ae,aecompl}
\usepackage{xcolor}

\usepackage{graphicx}	
\usepackage{amssymb}	
\usepackage{pdflscape}	
\usepackage[english]{babel}

\newcommand{\msol}{\mathrm{M}_{\rm \odot}}
\newcommand{\mjup}{\mathrm{M}_{\rm Jup}}
\newcommand{\mearth}{\mathrm{M}_{\rm \oplus}}
\newcommand{\mcore}{M_{\rm core}}
\newcommand{\mdotcore}{\dot{M}_{\rm core}}
\newcommand{\rcore}{R_{\rm core}}
\newcommand{\kms}{\mathrm{km \, s^{-1}}}

\newcommand{\tgap}{t_{\rm gap}}
\newcommand{\Cgap}{C_{\rm gap}}
\newcommand{\tmig}{t_{\rm mig}}
\newcommand{\Cmig}{C_{\rm mig}}
\newcommand{\tcross}{t_{\rm cross}}

\newcommand{\Cspace}{C_{\rm space}}

\title[Population synthesis of interacting disc fragments]{Towards a population synthesis model of self-gravitating disc fragmentation and tidal downsizing II: The effect of fragment-fragment interactions}

\author[D. H. Forgan, C. Hall , F. Meru and W.K.M. Rice]
{D.~H.~Forgan$^{1}$\thanks{Contact e-mail: \href{mailto:dhf3@st-andrews.ac.uk}{dhf3@st-andrews.ac.uk}},
C. Hall$^{2,3}$, F.~Meru$^4$, W.~K.~M.~Rice$^{2,5}$
\vspace{0.2cm} \\
$^{1}$Centre for Exoplanet Science, SUPA, School of Physics \& Astronomy, University of St Andrews, St Andrews KY16 9SS, UK \\
$^{2}$SUPA, Institute for Astronomy, University of Edinburgh, Blackford Hill, Edinburgh EH9 3HJ, UK \\
$^{3}$Department of Physics and Astronomy, University of Leicester, Leicester LE1 7RH, UK \\
$^{4}$Institute of Astronomy, Madingley Road, Cambridge CB3 0HA, UK\\
$^{5}$Centre for Exoplanet Science, University of Edinburgh, Edinburgh, UK}

\date{Accepted XXX. Received XXX; in original form XXX}

\pubyear{2017}

\begin{document}
\label{firstpage}
\pagerange{\pageref{firstpage}--\pageref{lastpage}}
\maketitle

\begin{abstract}
\noindent It is likely that most protostellar systems undergo a brief phase where the protostellar disc is self-gravitating.  If these discs are prone to fragmentation, then they are able to rapidly form objects that are initially of several Jupiter masses and larger.  The fate of these disc fragments (and the fate of planetary bodies formed afterwards via core accretion) depends sensitively not only on the fragment's interaction with the disc, but with its neighbouring fragments.  

We return to and revise our population synthesis model of self-gravitating disc fragmentation and tidal downsizing.  Amongst other improvements, the model now directly incorporates fragment-fragment interactions while the disc is still present.  We find that fragment-fragment scattering dominates the orbital evolution, even when we enforce rapid migration and inefficient gap formation.  

Compared to our previous model, we see a small increase in the number of terrestrial-type objects being formed, although their survival under tidal evolution is at best unclear.  We also see evidence for disrupted fragments with evolved grain populations - this is circumstantial evidence for the formation of planetesimal belts, a phenomenon not seen in runs where fragment-fragment interactions are ignored.  

In spite of intense dynamical evolution, our population is dominated by massive giant planets and brown dwarfs at large semimajor axis, which  direct imaging surveys should, but only rarely, detect.  Finally, disc fragmentation is shown to be an efficient manufacturer of free floating planetary mass objects, and the typical multiplicity of systems formed via gravitational instability will be low.

\end{abstract}




\section{Introduction}
\label{sec:introduction}




During the earliest phases of star formation, angular momentum conservation ensures that protostars form with a protostellar disc, with a mass typically comparable to that of the protostar \citep{Bate2010,Tsukamoto2016}.  The disc can therefore become gravitationally unstable if the Toomre parameter \citep{Toomre_1964}:

\begin{equation}
Q = \frac{c_s \kappa_{\rm ep}}{\pi G \Sigma} \sim 1,
\end{equation}

\noindent where $c_s$ is the sound speed, $\Sigma$ is the disc surface mass density and $\kappa_{\rm ep}$ is the epicyclic frequency (which in Keplerian discs is equal to $\Omega$).  When $Q$ is sufficiently low ($Q\sim 1.5-1.7$), the disc becomes unstable to non-axisymmetric perturbations \citep{durisen_review, Helled2014,Rice2016,Kratter2016}.  These perturbations grow into spiral density waves thanks to the disc's differential rotation, which can boost the local entropy by weak shock heating.  This increases $Q$, tending to push the system out of the instability regime (weakening the shocks).  If the disc is able to cool efficiently, the heating and cooling processes can enter an approximate balance, and the disc maintains a marginally stable state where $Q$ is self-regulated to be near unity \citep{Paczynski1978}.  In this balanced state, we can link the cooling rate through the dimensionless cooling time parameter $\beta_c$ \citep{Gammie}:

\begin{equation}
\beta_c = t_{\rm cool} \Omega^{-1},
\end{equation}

\noindent to the stress induced in the disc by the gravitational instability.  The stress induces a turbulent state (known as gravito-turbulence), meaning that the stress can be expressed as a turbulent pseudo-viscosity.  If we use the $\alpha$-parametrisation \citet{Shakura_Sunyaev_73}, then

\begin{equation}
\nu  = \alpha c_s H,
\end{equation}

\noindent where $H$ is the disc scale height.  If the disc is marginally stable and in local thermodynamic equilibrium,

\begin{equation}
\alpha = \left(\frac{d \ln \Omega}{d \ln r}\right)^{-2}\frac{1}{\gamma(\gamma-1)\beta_c}. \label{eq:alpha}
\end{equation}

\noindent In 2D simulations of gravito-turbulence, \citet{Gammie} showed that fragmentation occurred for $\beta_c \lesssim 3$, with the two dimensional ratio of specific heats $\gamma_{\rm 2D} =2$, or equivalently a minimum stress $\alpha \gtrsim 0.06$.  In 3D, \citet{Rice_et_al_05} confirmed that this minimum $\alpha$ criterion for fragmentation holds for various values of $\gamma$, which is commonly interpreted as the gravito-turbulent stress saturating at this approximate value.  Discs that can cool rapidly enough do not receive sufficient feedback from stress heating to prevent local condensations of gas collapsing under gravity.  In irradiated discs, a similar effect occurs if mass loading from the envelope is sufficiently rapid and does not result in strong accretion heating from shocks \citep{Kratter2011}.  Equivalently, if the local Jeans mass inside a spiral perturbation is able to decrease rapidly, then these perturbations are susceptible to prompt fragmentation \citep{Forgan2011a,Forgan2013}.

From a range of analytical calculations and numerical experiments, it is clear that cool, massive, extended discs will fragment promptly, i.e. within a few outer rotation periods of appropriate conditions being met \citep{Gammie,Rice2011,Mejia_2,BDNL,Rice_et_al_05,Cossins2008,Stamatellos2009,Stamatellos2011a,Steiman-Cameron2013, Backus2016}, and equally mass loading of a cool disc eventually results in fragmentation (see e.g. \citealt{collapses}).  What is less clear is the precise value at which gravito-turbulent stresses saturate in 3D, as earlier SPH simulations of self-gravitating discs have been shown to have issues with convergence \citep{Meru2011,Meru2012}.  A number of possible reasons (and algorithmic solutions) exist for this convergence issue \citep{Lodato2011,Rice2012,Rice2014}.  The previous five years of work in this area, which include particle-based and grid-based simulations suggest the critical $\alpha$ is unlikely to greatly exceed 0.1.  Most recently, meshless hydrodynamic simulations have demonstrated convergence at $\alpha \approx 0.13$, confirming that numerical dissipation is the cause \citep{Deng2017}.

It may still be the case that discs can fragment at lower stresses, if the power spectrum of density fluctuations in the disc occasionally permits very large over-densities \citep[so-called ``stochastic fragmentation'', ][]{Paardekooper2012}.  This process requires a significant time interval to occur, and for the over-densities to contract enough to be able to weather the subsequent spiral density waves \citep{Young2016}.

Fragmentation produces gaseous embryos with initial masses typically greater than a few Jupiter masses \citep{Rafikov2005,Boley2010b,Forgan2011a,Rogers2012}, with initial semimajor axes typically greater than 30 au.  These embryos are initially a sampling of disc material at the formation location, mostly gas but containing a population of dust grains.  

The fate of this dusty gas embryo is the subject matter of a revised model of the gravitational instability (GI) theory of planet formation, often referred to as ``tidal downsizing'' theory \citep[see e.g.][for a review]{Nayakshin2017}.  In short, a combination of several physical processes sculpts the embryo into one of a large number of final configurations.

The evolution of the embryo's gas envelope will be similar to that of the first-core/second-core evolution in protostars \citep{Masunaga_1}.  The grains in the embryo will continue to grow by collisions.  We expect this process to be slightly faster than in the surrounding disc as the density of gas is substantially higher, and the relative velocities between grains will be initially reduced.  

As the grains grow, they begin to feel an increasing drag force from the gas, pointed towards the pressure maximum at the centre of the embryo.  As the grain Stokes number tends towards unity, the grains begin to sink towards the centre\footnote{Note that in this context the Stokes number is defined as $S=t_{\rm stop} \Omega_{eddy}$, where the usual angular frequency $\Omega$ for discs is replaced with the fragment's eddy turnover frequency, which is determined either by the fragment's convective behaviour, or turbulence depending on which process dominates internal fluid motions.}.  This settling (opposed by turbulence and convection) increases the local density of dust, resulting in a mix of grain growth and grain fragmentation depending on the local grain velocity distribution.  This grain-heavy mixture at the centre of the embryo can undergo gravitational collapse, forming a solid core, which in itself can assist the final collapse of the gas \citep{Nayakshin2014a}.

During this period of internal evolution, the fragment is migrating inward.   Thanks to the significant torques generated by the self-gravitating disc, this migration can be rapid compared to the same process in non-self-gravitating discs \citep{Baruteau2011}.  Despite the fragment's relatively large mass, simulations indicate that driving a gap in self-gravitating discs is problematic, and hence the standard means by which migration is slowed is not always available \citep{Malik2015a}, although simulations also suggest the effect of radiative feedback from gas accretion can alter the migration state \citep{Stamatellos2015}.  The initial fragment radius is rather large (the typical Hill Radius for a fragment at birth is of order a few AU), and hence the fragment quickly experiences Roche lobe overflow and mass loss.

The final end product of GI is therefore a contest of various timescales; the gaseous collapse timescale, the grain growth and settling timescales, core formation timescales, migration timescales and mass loss timescales.  Ordering these timescales differently results in a zoo of objects spanning several orders of magnitude in mass, from brown dwarfs down to terrestrial planets.

An important question to ask is: \emph{how frequently do objects of a certain type form, given our knowledge of the expected initial conditions?} Or more generally, \emph{what does current GI theory predict for the observed exoplanet and brown dwarf populations?} 

In the first paper of this series, \citet{TD_synthesis} attempted to answer these questions by developing the first self-consistent population synthesis model of fragmentation and tidal downsizing.  This model combined three model components - a self-gravitating disc model, generalised fragmentation criteria for these discs using the Jeans criterion (with a self-consistent initial fragment mass), and a fragment evolution model.  

Their population synthesis model simulated the evolution of over a million disc fragments.  In only one case did an object form with the properties of a terrestrial planet in the inner 5 au of a planetary system.  The vast majority of objects were of masses greater than 13 $\mjup$ at semimajor axes greater than 30 au.  These bodies were considered to be brown dwarfs, with a secondary population of giant planets (both with and without solid cores).  Between 40 and 50 per cent of all fragments formed were completely destroyed by tidal interactions as they moved too close to the star.  Similar results were found by \citet{Galvagni2013}, who utilised 3D hydrodynamical collapse calculations as part of their analysis, and \citet{Nayakshin2015}, whose population synthesis model has some of the most advanced grain microphysics to date.  Notably, both models operate on a single-fragment per star basis, unlike the multiple fragmenting systems normally seen in hydrodynamic simulations. 

The model of \citet{TD_synthesis} was able to generate multiple fragments per system, but was only able to evolve the system until the disc had dissipated.  To investigate the fate of fragments after the disc phase \citet{TD_dynamics} used the output from the population synthesis model as input for subsequent N-Body integration, for both isolated multi-fragment systems, and in the tidal potential of their birth cluster.  In both sets of integrations, the GI objects showed significant potential for scattering to high eccentricities and semi-major axes, as well as a relatively high ejection rate from the system to form free-floating planets and field brown dwarfs.

This combined analysis allows the first statistical predictions of GI to be made that incorporate both the fragmentation process, and the subsequent dynamical evolution of the fragments.  The two datasets (before and after $N$-Body integration) are now being tested against observational data, in particular direct imaging surveys of exoplanets and brown dwarfs on wide orbits \citep{Vigan2017}.

One important weakness of both the \citet{TD_synthesis} data and the \citet{TD_dynamics} data was an absence of fragment-fragment interaction \emph{while the disc is still present}.  The initial population synthesis runs evolved the fragments effectively in isolation - although they inhabited the same disc, they did not interact with each other and could not influence each other's early dynamical evolution.  

\citet{Hall2017} analysed several SPH realisations of a fragmenting disc, tracking the properties of the fragments over several thousand years of disc and fragment evolution. They retrieve fragment destruction rates nearly half of that predicted by \citet{TD_synthesis}.   The population synthesis model also fails to reproduce the semi-major axis distribution produced by the hydrodynamic simulations, and the mass-semimajor axis relationship is markedly different, with eccentricities of order 0.1 (compared to the circular orbits assumed by the population synthesis model).  

In short, the first several thousand years of a fragment's existence is largely governed by dynamical interactions, and any population synthesis model worth its salt must be able to account for this.  In this work, we present a significant upgrade to the model of \citet{TD_synthesis}.  Our model now implements direct N-Body integration of each star system, to capture both the migration of the fragments, and their interactions during the earliest phases of their existence.

Section \ref{sec:method} describes the advances we have made in population synthesis modelling of GI; section \ref{sec:results} shows the effects of the new physical processes on the resulting population.  Section \ref{sec:discussion} notes future directions for model development, and implications of the model for both the bound and free floating populations of substellar objects, and in section \ref{sec:conclusions} we conclude the work.

\section{Method}
\label{sec:method}

\noindent For brevity, we only describe in detail the major changes to the population synthesis model initially presented in the first paper in this series \citep{TD_synthesis}, and refer the reader to this previous work for further information.

A series of 1D self-gravitating disc models including photoevaporation \citep{Rice_and_Armitage_09,Owen2011}, are run in advance.  These are the same models as used in \citet{TD_synthesis}. A total of 100 models are used, with the protostellar mass varying between $0.8$ and $1.2$ $\msol$\footnote{This is quite a limited range, but it does allow a certain control over the disc's thermodynamic properties.  Tests with more massive stars show quite similar behaviour (with more massive fragments being a typical result)}.

Each disc has a maximum extent of 100 au, and a surface density profile $\Sigma \propto r^{-1}$.  The total disc mass is selected so that the disc-to-star mass ratio varies between 0.125 and 0.375.  We select this range as the disc is unlikely to be self-gravitating below a mass ratio of 0.1, and discs with mass ratios above 0.4-0.5 rapidly accrete onto the star \citep{Forgan2011}.

The discs have a fixed value of $Q$, and are evolved viscously, where the local value of $\alpha$ is determined from the cooling rate according to equation (\ref{eq:alpha}).  The cooling rate depends on the stellar irradiation and the local optical depth, which is computed using the opacity tables of \citet{Bell_and_Lin}.

We evolve each disc for 1 Myr (or until the disc dissipates), which depends on the strength of the X-Ray luminosity (which is also randomly sampled between $5\times 10^{28}$ and $10^{31} \mathrm{erg \, s^{-1}}$).  The disc's evolution is stored as a series of snapshots, taken every 1000 years.  As the disc evolution timescale is typically much larger than the fragment evolution timescale, the disc's properties (mainly surface density and sound speed) are linearly interpolated in time between  snapshots.

\begin{figure}
\begin{center}
\includegraphics[scale=0.35]{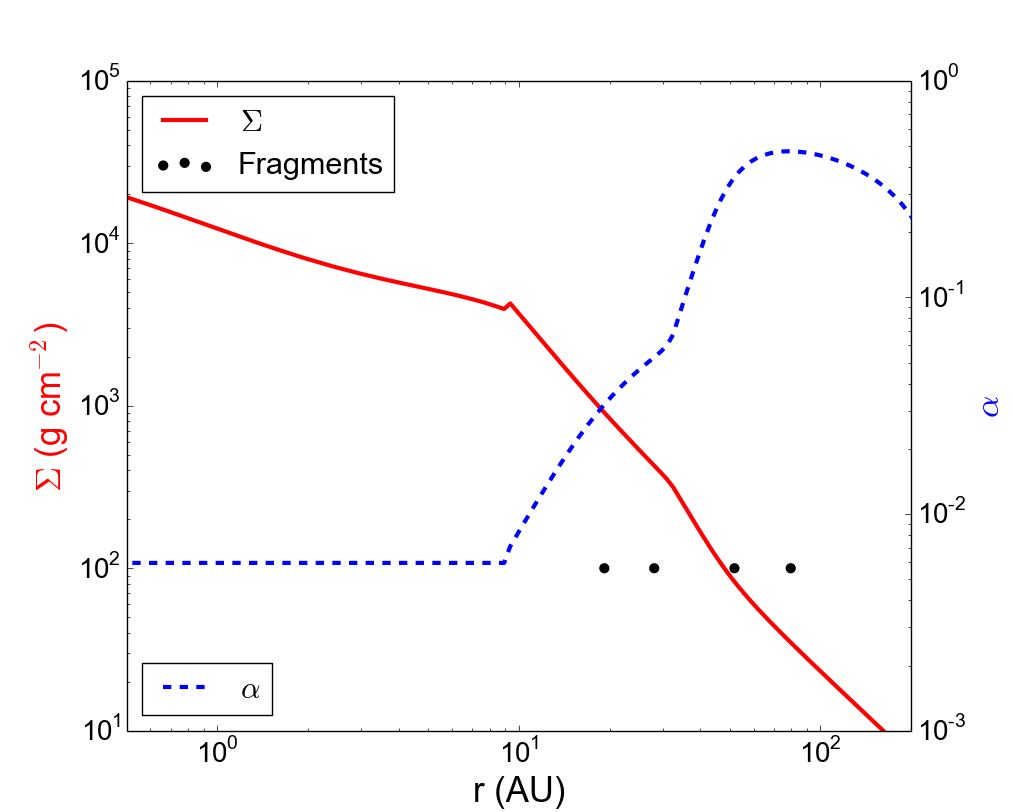}
\end{center}
\caption{The initial surface density profile (red solid line) and viscous transport parameter $\alpha$ (dashed blue line) near the beginning of a model run.  The circles indicate the initial radial location of fragments in this disc. \label{fig:discprofile}}
\end{figure}

When a system is simulated in the population synthesis model, a disc model is selected.  Fragments are added to this disc at time $t=0$ by calculating the smallest radius at which fragmentation will occur \citep{Forgan2011a}.  Fragments are then assigned with an initially random spacing between $[1.5-\Cspace]$ Hill Radii.  In \citet{TD_synthesis} we fixed $\Cspace=3$, as the initial fragment spacing had little effect on individual fragment outcomes.  In this work, the initial fragment spacing will govern the initial strength of fragment interactions, so we will investigate this parameter's influence.  An example of a surface density profile and an effective viscous alpha profile, near the beginning of a model run (and fragments placed therein), is shown in Figure  \ref{fig:discprofile}.

Once fragments are seeded into the disc, their internal evolution is governed by the system of equations elucidated by \citet{Nayakshin2010,Nayakshin2010a,Nayakshin2010b}, and fully described in \citet{TD_synthesis}. 

\subsection{Revised migration model}

\noindent Originally, the migration implementation was selected for consistency with Nayakshin's prescription, so that the outputs of their complete system of tidal downsizing equations could be interrogated on a statistical level (for a limited range of free parameters).  In this system, the criteria for the transition between Type I and Type II migration (equivalently, the criteria for opening a gap) depended only on the fragment mass and aspect ratio $H/R$.  

In reality, the strength of viscous torques also plays a role. To this end we now utilise the torque-balance criterion of \citet{Crida2006} to determine gap opening:

\begin{equation}
\frac{3}{4}\frac{H}{R_H} + \frac{50 \nu}{q \Omega_p R^2_p} \leq 1,
\end{equation}

\noindent where $\nu= \alpha c_s H$ as usual, the fragment to star mass ratio is $q=M_p/M_*$, and the Hill Radius

\begin{equation}
R_H = a_p \left(\frac{q}{3}\right)^{1/3}.
\end{equation}

\noindent Following \citet{Malik2015a}, we also demand that the gap opening time be less than the gap crossing time, $\tgap<\tcross$.  The gap crossing time is estimated assuming that the half-width of the horseshoe region is approximately $2.5 R_H$:

\begin{equation}
\tcross = \frac{2.5 R_H}{v_{\rm mig}}.
\end{equation}

\noindent We estimate $v_{\rm mig}$ by using

\begin{equation}
v_{\rm mig} = \frac{R}{\tmig},
\end{equation}

\noindent where $R$ is the distance to the central star, and $\tmig$ is the migration timescale assuming Type I migration (see below).  The gap opening timescale is \citep{Lin1986}:

\begin{equation}
\tgap = \Cgap \left(\frac{H}{R}\right)^5  \frac{1}{q^{2}\Omega},
\end{equation}

where $\Cgap$ is a free parameter which we set to unity unless specified.  If either the torque-balance criterion is not satisfied or $\tgap>\tcross$, then the fragment cannot open a gap and Type I migration is in effect:

\begin{equation}
\tmig = \Cmig \left(\frac{H}{R}\right) \frac{1}{q\Omega},
\end{equation}

\noindent where again $\Cmig$ is a free parameter.  If both the torque-balance criterion is satisfied and $\tgap<\tcross$, then the fragment migrates via Type II:

\begin{equation}
\tmig = \frac{2}{3}\frac{1}{\alpha \Omega} \left(\frac{H}{R}\right)^{-2},
\end{equation}

\noindent where we assume that the disc mass remains large enough to avoid the planet-dominated regime. Note that $\Cmig$ does not affect the migration timescale if Type II is in effect.

\subsection{Radiative Feedback from Core Formation}

\noindent \citet{Nayakshin2016} noted a new destruction mechanism for GI objects that form cores.  The accretion luminosity of the core 

\begin{equation}
L_{acc,core} = \frac{G \mcore \mdotcore}{\rcore},
\end{equation}

is potentially a very large source of energy.  This radiative feedback is dumped into the surrounding envelope, affecting its overall boundness.  If this luminosity is sufficiently large, core formation can produce radiative feedback strong enough to unbind the entire embryo.

This places an approximate upper core mass for GI objects of some tens of $\mearth$.  \citet{Nayakshin2016} argues that this process can destroy a great fraction of the more massive GI objects expected to reside at large separations, that remain largely absent from observations \citep{Bowler2014a,Vigan2017}.

Our simulations as yet do not include gas or solids accretion after formation (see Discussion), so we cannot fully assess the effects of this process.  We can however make a simple estimate by comparing the total energy released by core formation to the binding energy of the embryo, and derive the condition

\begin{equation}
\frac{\mcore^2}{\rcore} > \frac{M_p^2}{R_p}.
\end{equation}

\noindent For destruction by radiative feedback.  We find that for all runs, no fragments satisfied this criterion for core formation mainly due to low core masses.  This confirms that this feedback process is only effective if the embryos accrete significant quantities of solids.

\subsection{Initial Fragment Locations}

\noindent In \citet{TD_synthesis}, the initial positions of fragments were determined by identifying the fragmentation boundary, and placing a fragment there with mass $M=M_J$, where $M_J$ is the local Jeans mass inside a spiral perturbation \citep{Forgan2011a}.

The semimajor axis of the next fragment is then given by

\begin{equation}
a_{\rm next} = a_{\rm previous} + (1.5 + \eta(\Cspace-1.5))R_H(M_J),
\end{equation}

\noindent where $R_H(M_J)$ is the Hill Radius of the previous fragment, $\eta$ is a uniformly sampled random number in the range $[0:1]$, and $\Cspace$ is a free parameter.  More simply, we ensure that fragment spacings vary uniformly between 1.5 and $\Cspace$ Hill Radii.  This results in a maximum number of 5 fragments for our disc model parameters, with 3 being the typical initial fragment multiplicity.

In all previous runs of the model, $\Cspace$ was fixed at 3, and as fragment-fragment interactions were not previously modelled during the disc phase, changing $\Cspace$ had very little effect on the resulting population.

In the new version of the model, $\Cspace$ is now a key factor in determining the strength of fragment-fragment interactions in the earliest disc phase (see following sections).  Fragment spacings can also be modified by migration, so it is important to determine whether spacing or migration determines the resulting population.  In section \ref{sec:results} we will show how varying $\Cspace$ now has significant effects on the resulting population.

\subsection{Fragment-Fragment Interactions}

\noindent The population synthesis model can now run in one of two modes.  In the first, the orbital evolution of the fragments proceeds as in \citet{TD_synthesis}, where each fragment's semimajor axis is decreased accordingly to the local value of $\tmig$.  As a result, the fragments' orbits remain circular with zero inclination.

In the second, the fragment's orbits are evolved in 3D via N-Body integration.  We use the prescription described in \citet{Alibert2013}, which calculates the gravitational force on embryo $i$ due to the other embryos $j$ in the heliocentric frame, i.e. we fix the star at the origin.  The gravitational acceleration experienced by $i$ is

\begin{equation}
\mathbf{\ddot{r}}_{i} = -G (M_* + M_i)\frac{\mathbf{r}_{i}}{\left|\mathbf{r}_i\right|^3} - G \sum_{j=1, j \neq i}^{N_{\rm embryo}} m_j \left( \frac{\mathbf{r}_{i} - \mathbf{r}_{j}}{\left|\mathbf{r}_{i} - \mathbf{r}_{j}\right|^3} + \frac{\mathbf{r}_{j}}{\left|\mathbf{r}_{j}\right|^3} \right).
\end{equation}

\noindent In the above, $\mathbf{r}_i$ is the position vector of body $i$ relative to the star.  Note that these dynamical interactions ignore tidal forces, and the rotation of all bodies.  Radial migration is modelled via the following drag term \citep{Fogg2007}:

\begin{equation}
\mathbf{a}_{\rm mig,i} = - \frac{\mathbf{v}_{i}}{2 \tmig}.
\end{equation}

\noindent Note that while the fragment motions are 3D, the migration timescale is defined by the azimuthally symmetric 1D disc properties.  The orbital eccentricity and inclination of the embryos should also be damped by their interactions with the disc.  To implement this we add the following extra drag terms:

\begin{equation}
\mathbf{a}_{\rm damp,e} = - 2\frac{\left(\mathbf{v}_{i}.\mathbf{r}_{i}\right)\mathbf{r}_{i}}{\left|\mathbf{r}_{i}\right|^2 \chi \tmig} \label{eq:eccdamp}
\end{equation}

\begin{equation}
\mathbf{a}_{\rm damp,i} = - 2\frac{\left(\mathbf{v}_{i}.\mathbf{\hat{z}}\right)\mathbf{\hat{z}}}{\chi \tmig}. \label{eq:incdamp}
\end{equation}

\noindent We use the $\chi$ parameter to estimate both the eccentricity and inclination damping timescales in terms of $\tmig$.  Both damping timescales are certainly significantly longer than $\tmig$: we follow \citet{Alibert2013} by setting $\chi=10$\footnote{Note that in general there is no expectation for the eccentricity and inclination damping timescales to be the same.  As we will show, our fragments tend towards low inclination orbits, and the eccentricity evolution is dominated by scattering.}.  The system of equations is integrated via a 4th order Runge-Kutta formalism, with a typical adaptive timestep algorithm.  Tests of the algorithm on undamped N Body systems show that energy is conserved to better than one part in $10^5$ over the entire integration for typical initial fragment masses and positions.

\section{Results}
\label{sec:results}

\noindent We now explore the role of fragment-fragment interactions by running the model in seven different configurations.  These include two control runs where $N$-body interactions are switched off and on respectively, and a further five runs modifying the initial fragment separation ($\Cspace$), the initial fragment eccentricity and inclination, and the migration parameters $\Cmig$ and $\chi$.  Table \ref{tab:run-params} lists all the runs described in this work, with the parameters used for each.

For each model run, we produce approximately 30,000 individual planetary systems, with a total of at least 100,000 fragments initially.  With destruction rates being of order 40\%, this results in around 50,000-60,000 final objects (both bound and free-floating).

\begin{table}
\caption{The model runs in this paper, with the values of each parameter for each run. The control runs C-off and C-on are identical, with the exception of N-Body physics off or on respectively.\label{tab:run-params}}
\begin{tabular}{ccccccc}
\hline

Run & N-Body & initial $e$ & $\Cspace$ & $\Cmig$ & $\Cgap$ & $\chi$ 		\\
\hline
C-off/C-on & No/Yes & 0    &  3  & 1   & 1           & 10 \\
W      & Yes & 0   & 10 & 1   & 1           & 10 \\
E       & Yes & >0 & 3  & 1    & 1           & 10 \\
R       & Yes & 0   & 3  & 0.1 & 1           & 10 \\
G       & Yes & 0   & 3  & 1    & $10^5$ & 10 \\
RG     & Yes & 0   & 3  & 0.1 & $10^5$ & 10 \\
RGE   & Yes & 0   & 3  & 0.1 & $10^5$ & 1 \\
\hline
\end{tabular}
\end{table}

\subsection{Control Runs - Switching N Body physics on and off}

\begin{figure*}
\begin{center}$\begin{array}{cc}
\includegraphics[scale=0.4]{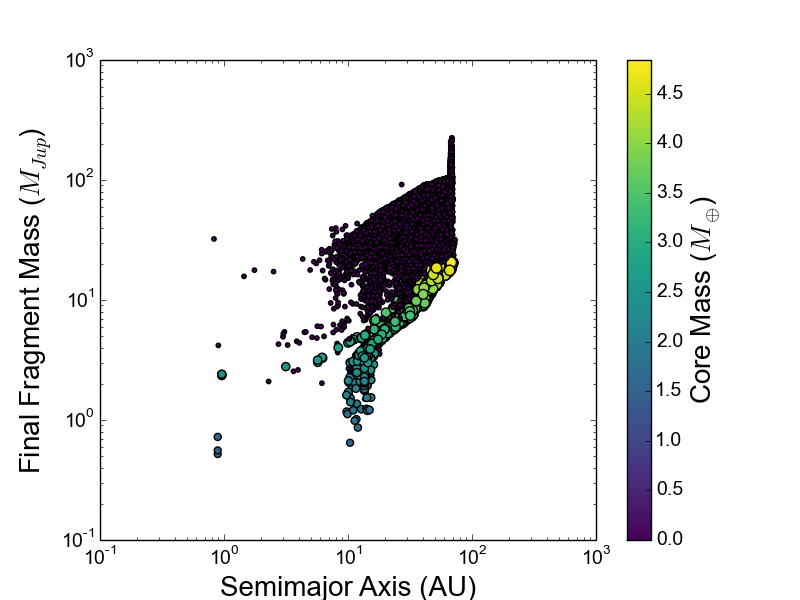} &
\includegraphics[scale=0.4]{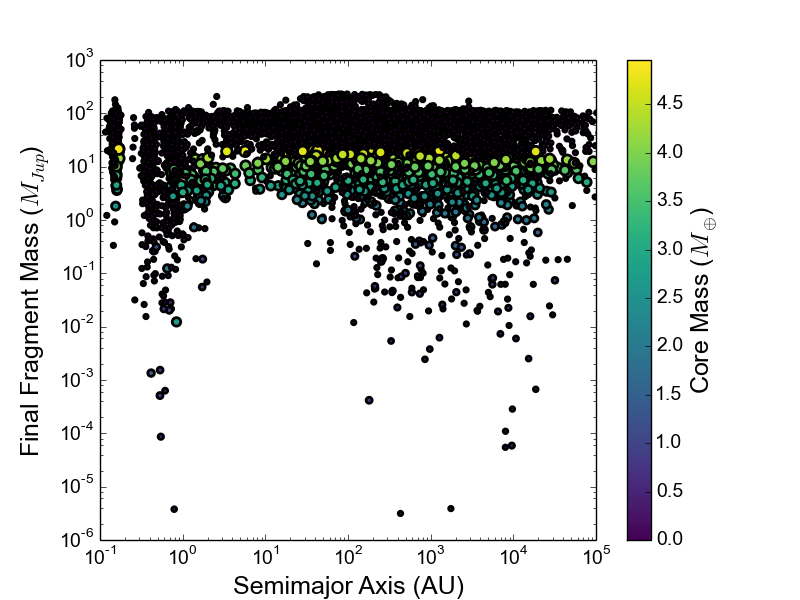} \\
\end{array}$
\end{center}
\caption{Investigating the role of fragment-fragment interactions.  Left: The population derived with N-Body physics switched off (run C-off).  Right: the same model run with N-Body physics active (run C-on, $\Cspace=3$).  In both cases, many of the resulting bodies remain close to their birth locations of $a=30-100$ au.  In the absence of fragment-fragment interactions (left), orbit modification is caused by migration only.  Fragments which can successfully open a gap migrate more slowly, producing the stream of $\sim \mjup$ bodies with substantial cores (as well as the brown dwarf population).  With interactions active (right), scattering causes significant orbit modification, resulting in clusters of low mass bodies at $a\sim$ 0.1 and 1 au (with eccentricities >0.5) due to inward scattering, and beyond 100 au (with lower eccentricities) due to outward scattering.  More massive bodies sacrifice angular momentum to eject neighbouring fragments, and hence occupy the 1-10 au region while avoiding catastrophic migration and destruction \label{fig:control}}
\end{figure*}

\begin{figure}
\begin{center}
\includegraphics[scale=0.4]{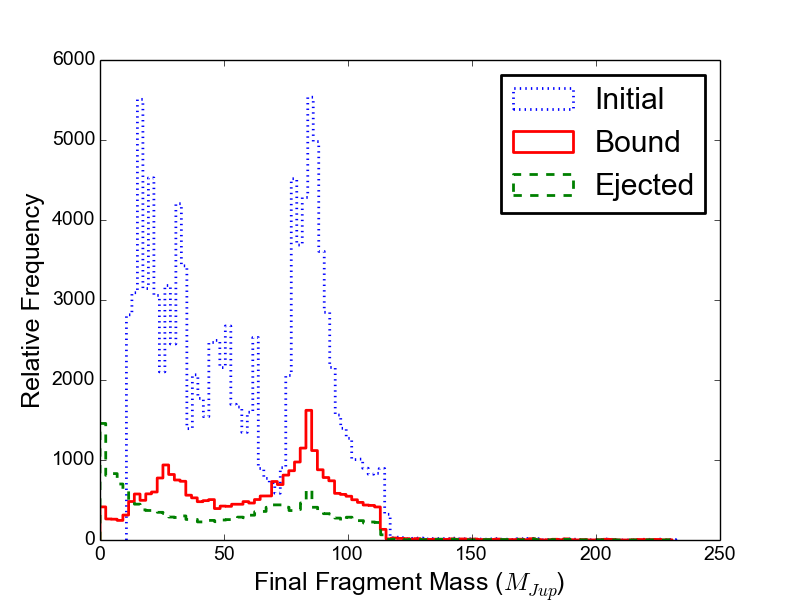}
\end{center}
\caption{The mass distribution of all objects initially (blue dotted line), and the surviving bound (red, solid) and ejected (green, dashed) objects for run C-on. \label{fig:mhist_compare}}
\end{figure}

\begin{figure*}
\begin{center}$\begin{array}{cc}
\includegraphics[scale=0.4]{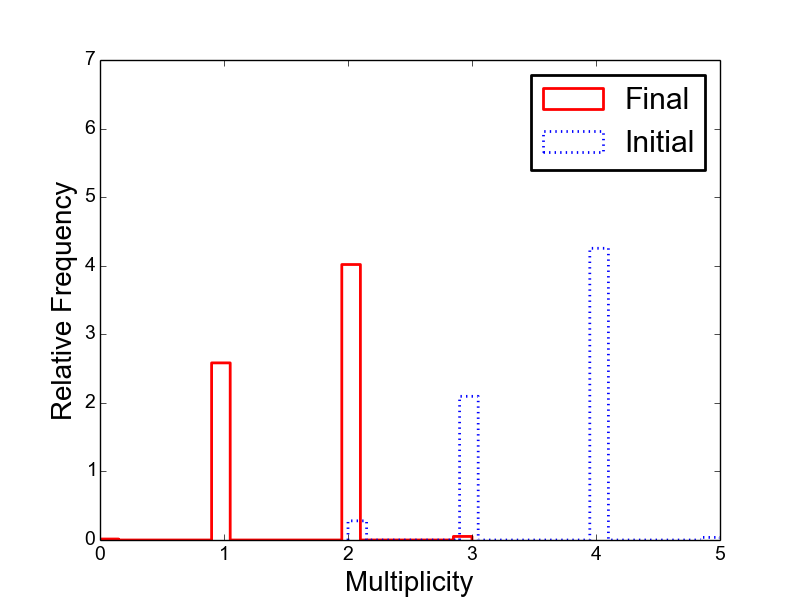} &
\includegraphics[scale=0.4]{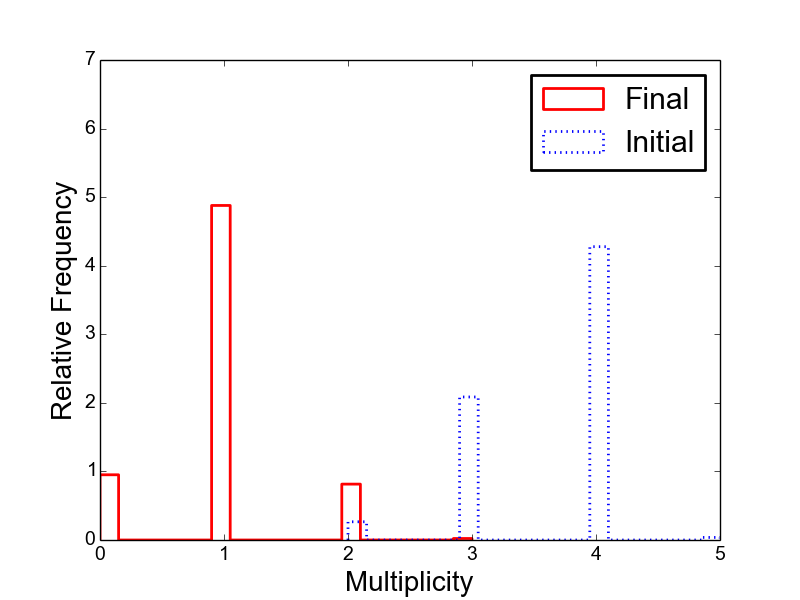} \\
\end{array}$
\end{center}
\caption{The multiplicity statistics of systems initially at $t=0$ and  at the end of the simulation, i.e. the final number of fragments still bound to the star, with N-Body physics off (run C-off, left), and on (run C-on, right).  \label{fig:control_mult}}
\end{figure*}

\begin{figure*}
\begin{center}$\begin{array}{cc}
\includegraphics[scale=0.4]{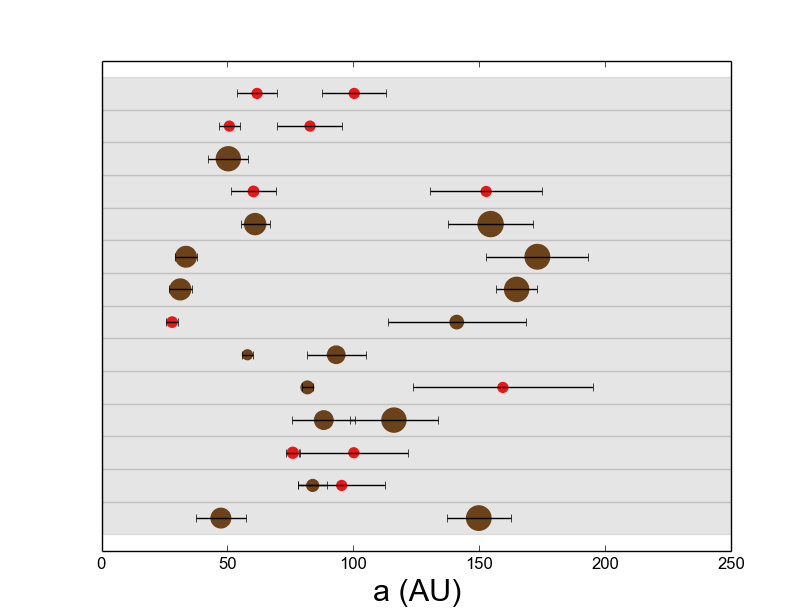} &
\includegraphics[scale=0.4]{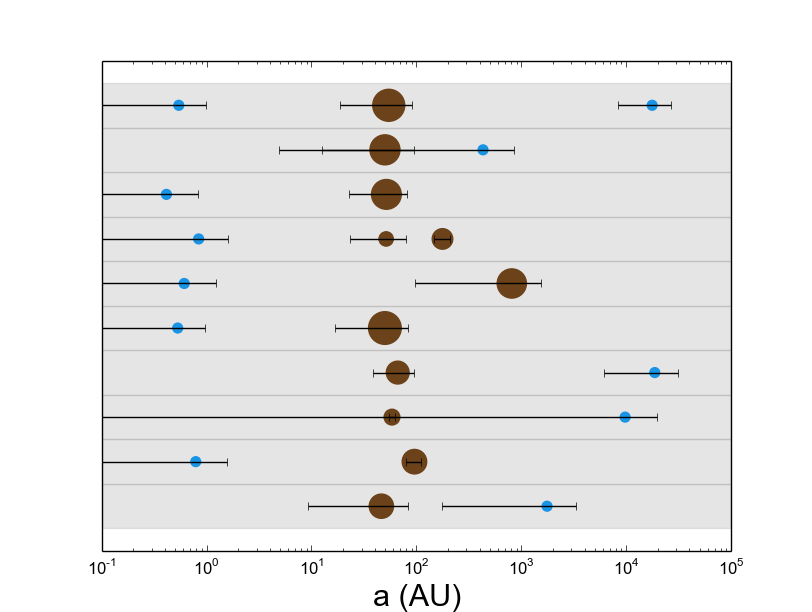} \\
\end{array}$
\end{center}
\caption{Portraits of multibody systems after 1 Myr of evolution, produced in the control run with N-Body physics active (run C-on). Each row represents an individual system.  Brown circles indicate brown dwarfs, which we define as bodies with masses above the canonical limit of $13 \mjup$.  Red circles indicate gas giant planets (both with and without cores).  Blue circles indicate rocky planets (where the core mass is more than 50\% of the total mass). The sizes of brown and red circles indicate their relative masses (blue circles are not to scale). The error bars indicate the apastron and periastron of the body's orbit.  Left: The most common types of stable multiple planet systems, with bodies greater than several Jupiter masses.  Right: The least common types of system are those in which terrestrial planets are present (usually accompanied by a more massive giant planet or brown dwarf).  Terrestrial planets compose less than 0.05\% of the entire surviving population still bound to the star. \label{fig:control_portraits}}
\end{figure*}

\noindent To begin with, we run two control models, where we keep all parameters fixed, but switch between simple orbital migration without N-Body effects, and the full N-Body integration formalism as described in this paper (runs C-off and C-on respectively).  The migration and gap parameters $\Cmig=\Cgap=1$, and $\Cspace=3$.  All fragments are formed with zero eccentricity and inclination.

Figure \ref{fig:control} shows the mass vs semimajor axis space for both runs (the size and colour of each point both indicate the core mass in Earth masses, as shown by the colour bar - black points are core-less objects).

With N-Body physics switched off, the population looks similar to that produced by \citet{TD_synthesis}.  The changes to the calculation of $\tmig$ (using the torque-balance plus gap-opening criteria) result in slightly more efficient migration.  The core masses as a result tend to be restricted to lower masses, and fewer giant planets with cores of a few Earth masses arrive at distances of a few AU than the equivalent run with the previous model (Figure 10 of \citealt{TD_synthesis}).  

Rerunning the model using the N-Body integrator has a profound effect on the resulting population.  While the overall destruction rate is around 40\% whether N-Body physics is active or otherwise, fragment-fragment scattering spreads the bodies throughout the available semi-major axis space, from the inner simulation boundary at 0.1 AU to several mega-AU (i.e. tens of pc), with a reduction of bodies in the 1-15 AU range at masses above a few Jupiter masses (correspondingly roughly to the Brown Dwarf Desert).  As these systems are likely to still be in their birth environment during this phase, bodies orbiting at greater than a few thousands of AU are likely to be liberated from the system by encounters with nearby stars and become free-floating bodies \citep{TD_dynamics}.

Even without a perturbing cluster potential (or Galactic tides), approximately 38\% of the fragments that are not destroyed are ejected from the system (i.e. around 15\% of all fragments formed), with velocities at infinite separation peaking at  4-5 $\kms$.  Figure \ref{fig:mhist_compare} shows the mass distribution for all fragments initially, as well as the final distribution for bound objects and ejecta.  The Jeans mass increases monotonically from distance to the star, with the minimum allowed mass determined by the disc's inner fragmentation boundary of around 30 au, and the maximum allowed mass found at the disc's outer boundary of 100 au.  The value of $\Cspace$ governs where the next fragment may appear (and how many fragments are possible to fit into the disc).  As a result, the initial mass distribution shows two strong peaks at around 10 and 100 $\mjup$ (with a smaller third peak around 50 $\mjup$).  Fragment evolution and dynamical processing preserves a two-peak mass function, with low mass objects tending to be destroyed due to their proximity to the star (and faster migration rates), and more massive objects being sufficiently distant to avoid destruction.  Intermediate mass objects are scattered and are either destroyed or preserved depending on the scattering event.  It is worth noting that this mass function depends explicitly on the disc's mass and outer radius, and therefore we should expect this two-peak distribution to broaden as we consider a wider range of disc properties.

Given that ejections are not possible with N-Body integration switched off, it is obvious that the multiplicity of planetary systems produced by the population synthesis model will change (Figure \ref{fig:control_mult}).  Not only does scattering reduce the number of systems with 2 or more orbiting bodies, it also increases the number of systems with no surviving bodies.  Typically, systems with 2 or more bodies will undergo at least one scattering event that results in ejection.  The resulting angular momentum exchange sends one fragment into the inner regions of the system, resulting in tidal disruption proceeding at even greater efficiency than would otherwise be the case.  This tends to remove pairs of fragments - one via ejection, one via disruption.  The left panel of Figure \ref{fig:control_portraits} shows typical examples of stable multiple systems, containing giant planets or brown dwarfs well in excess of $1\mjup$.

Of the surviving bodies still bound to the star, we can see that scattering allows some objects to form cores and lose most of their envelopes, forming terrestrial type planets of a few $\mearth$.  The right panel of Figure \ref{fig:control_portraits} shows the systems these rare objects tend to inhabit. We define a body as ``terrestrial'' if its core mass is greater than 50\% of the total body mass.  This category therefore includes super-Earths, but typically excludes mini-Neptunes. 

A note of caution is necessary here, as these terrestrial bodies often have low semimajor axes and high eccentricities, and are likely to either undergo tidal evolution to reduce both their eccentricity and semi-major axis, or to indeed plunge into the central star.  As such, their survival as warm rocky bodies is not guaranteed.  

Of the destroyed bodies, we note in a very small number of cases that the solids component had successfully completed grain growth and was in the process of sedimenting to form a core.  Once disrupted, such bodies would potentially have produced planetesimal belts at the disruption radius, depending on the dynamical circumstances \citep{Nayakshin2012}.  This was never seen in the original \citet{TD_synthesis} model runs, and seems to be possible only as a result of fragment-fragment interactions permitting fragments extra time to evolve before reaching their Roche limit.

We should also note that in reality, the more massive bodies in these simulations are likely to make a significant dynamical effect on the central star, resulting in strong changes in the system centre of mass.  Our $N$-body prescription is heliocentric, and in this sense the resulting dynamics of the bodies correctly resolve these effects, but the disc model does not.  This is an issue that requires resolution in future work (see Discussion).

\subsection{The Effect of Initial Fragment Separation}

\begin{figure}
\begin{center}
\includegraphics[scale=0.4]{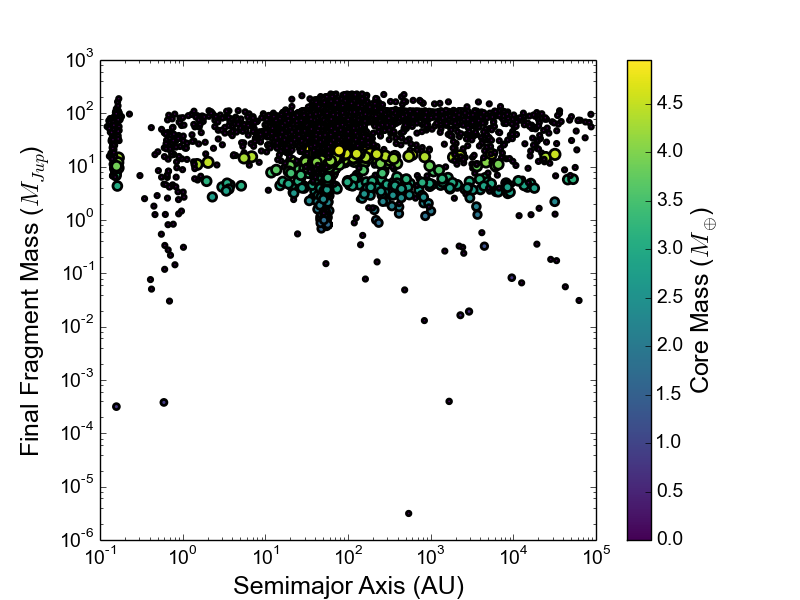}
\end{center}
\caption{Rerunning the control model with a larger $\Cspace=10$ (run W). \label{fig:cspace}}
\end{figure}

\begin{figure*}
\begin{center}$\begin{array}{cc}
\includegraphics[scale=0.4]{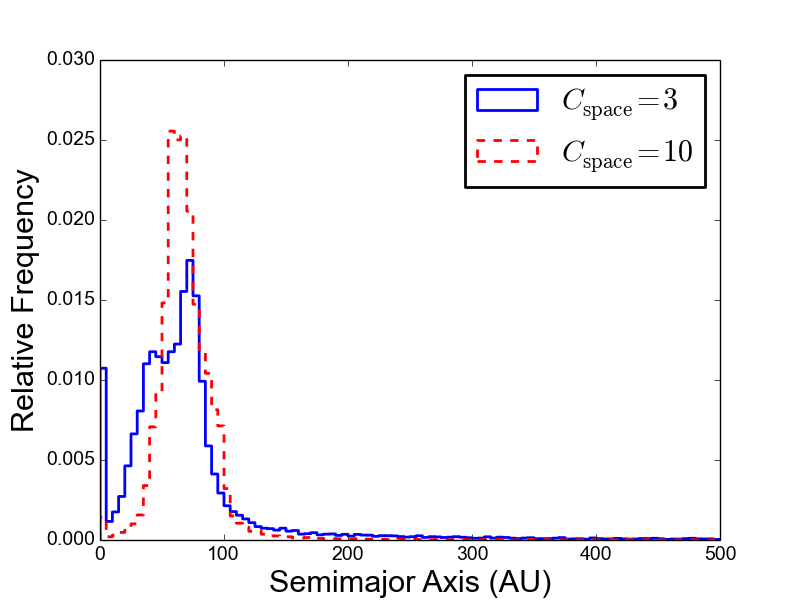} &
\includegraphics[scale=0.4]{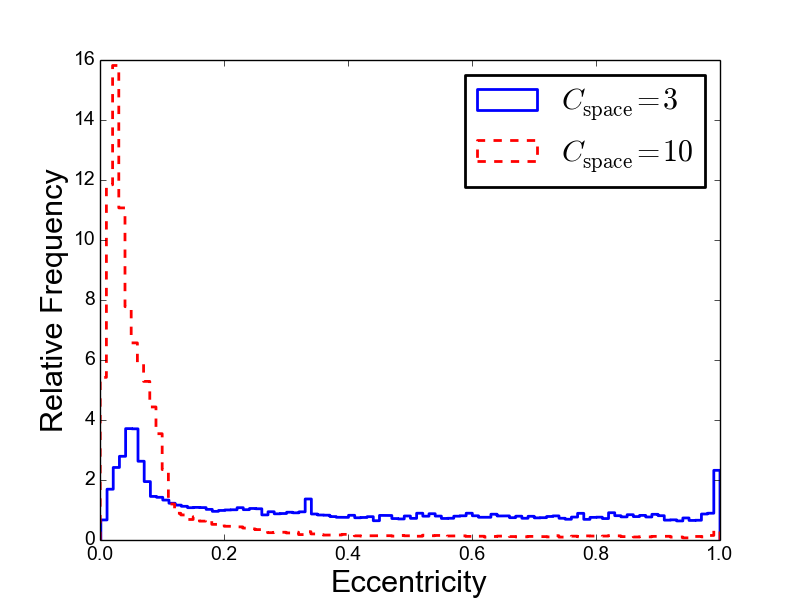} \\
\end{array}$
\end{center}
\caption{Left: The semimajor axis distribution of bound fragments for runs with $\Cspace=3,10$ (C-on and W respectively). Right: The eccentricity distribution.  \label{fig:cspace_ea}}
\end{figure*}

\begin{figure}
\begin{center}
\includegraphics[scale=0.4]{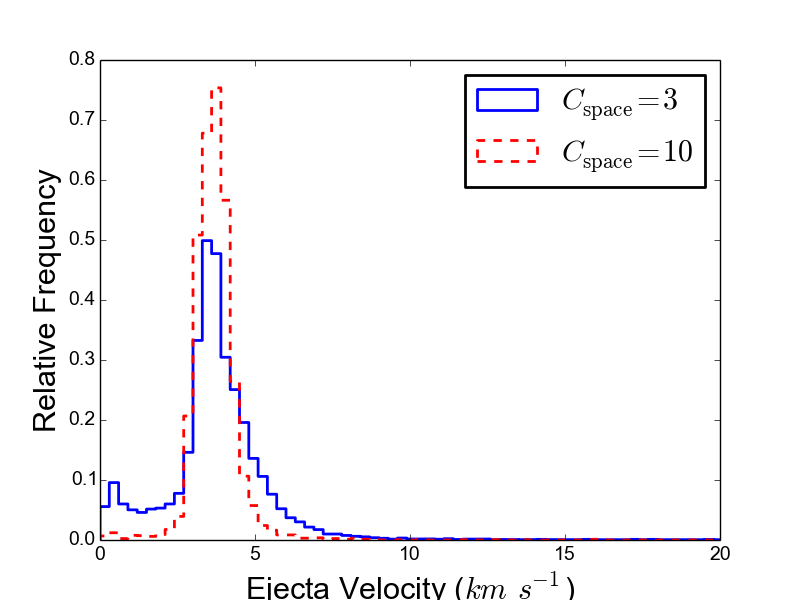}
\end{center}
\caption{The distribution of velocities for ejected bodies (measured at infinity), for runs with $\Cspace=3,10$ (C-on and W respectively). \label{fig:cspace_ejecta}}
\end{figure}

\noindent To test the strength of dynamical interactions, we vary the initial fragment spacing from its control value of $\Cspace=3$ to $\Cspace=10$ (run W, Figure \ref{fig:cspace}).  Increasing $\Cspace$ reduces the initial multiplicity of each system, reducing the number of potential scattering events.  The probability of two neighbouring bodies experiencing orbit crossing depends sensitively on their mutual Hill radius.  Bodies whose orbital separation is small compared to the mutual Hill radius (or equivalently $\Cspace$) are more likely to interact and produce orbit crossing events which lead to scattering.

It is therefore unsurprising to see that the $\Cspace=10$ run has far less scattering than the control run.  The semimajor axis and eccentricity distributions are both more strongly peaked, with most orbits remaining close to circular at semimajor axes beyond 50 AU  (Figure \ref{fig:cspace_ea}).

We see no evidence of migration resulting in resonant capture and convergent migration in any of the runs conducted in this paper, mainly because the migration rates depend strongly on the fragment mass, which is decreasing at a rate proportional to distance from the star.  The inner fragments lose mass more quickly, and hence begin to migrate inward more rapidly.  As a result, the inner fragments tend to move away from the outer fragments over time, forbidding resonant capture.

The destruction rate remains similar, but the ejection rate drops to around 9\%.  The ejecta velocities are also more tightly peaked around $4 \kms$ (Figure \ref{fig:cspace_ejecta}).

It is worth noting that this run adopts a relatively high $\Cspace$ compared to derived values from simulations, which suggest $\Cspace\sim 1$ \citep[see e.g.][]{Stamatellos2011a,Meru2015,Hall2017} and hence in reality higher amounts of scattering is expected.

\subsection{Varying the initial fragment eccentricity/inclination}

\noindent In our control run, the initial eccentricity and inclination of the fragments were both set to zero.  Using a suite of 9 3D hydrodynamic simulations of fragmenting discs, \citet{Hall2017} computed Gaussian fits for the initial distribution of both parameters, and as such we now attempt a run with these parameters (run E, see Table \ref{tab:ecc}).

\begin{table}
\centering
\caption{Gaussian fits for the initial eccentricity and inclination of fragments from \citet{Hall2017}. Eccentricities are constrained to be greater than zero by discarding any negative draws from the Gaussian variate, and inclinations are always converted to positive values. \label{tab:ecc}}
  \begin{tabular}{c || ccc}
  \hline
   Parameter & $\mu$ & $\sigma$   \\
   \hline
$e$ & 0.094 & 0.095 \\
$i (^\circ)$ & 0.00091 & 0.00005 \\
  \hline
\end{tabular}
\end{table}

\begin{figure}
\begin{center}
\includegraphics[scale=0.4]{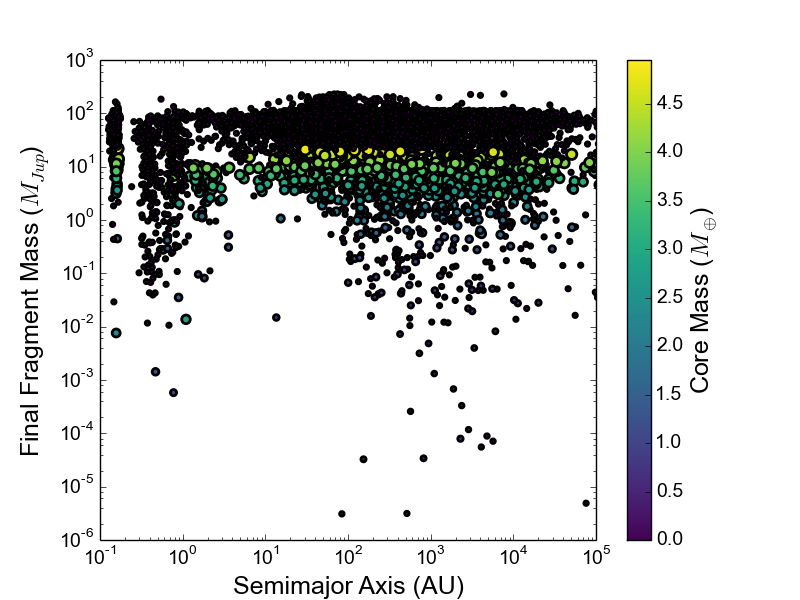}
\end{center}
\caption{Rerunning the control model with non-zero initial eccentricities and inclinations (run E).  The distributions of each can be found in Table \ref{tab:ecc}.\label{fig:initialecc}}
\end{figure}

Figure \ref{fig:initialecc} shows the resulting bound population produced.  There is no appreciable change in either the semimajor axis or eccentricity distribution of the objects bound to the star, which is unsurprising as the initial eccentricities and inclinations are negligible.

\begin{figure}
\begin{center}
\includegraphics[scale=0.4]{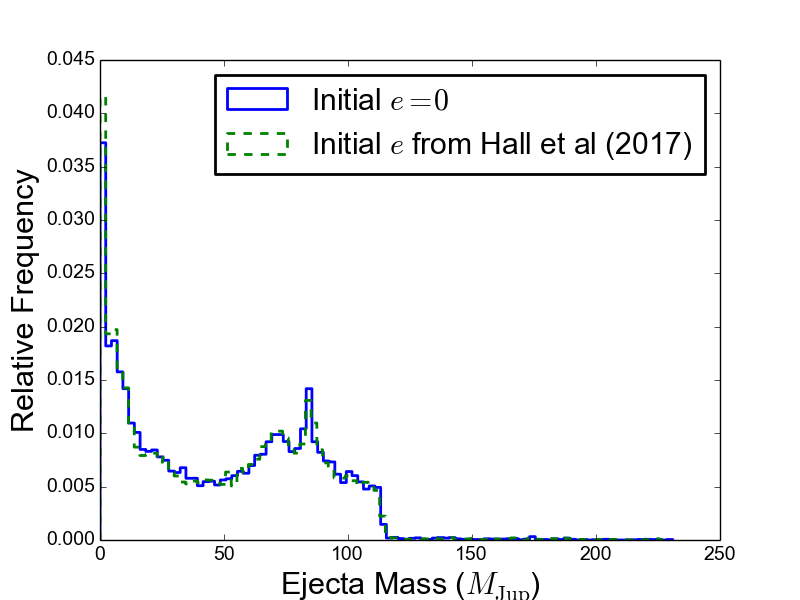}
\end{center}
\caption{The mass distribution of ejected bodies in the control run (zero initial eccentricity/inclination, run C-on), versus the run with initial eccentricities and inclinations taken from the simulations presented in \citet{Hall2017} (run E).\label{fig:initialecc_ejectamass}}
\end{figure}

The same is true for the properties of the ejected objects.  We plot the ejecta mass function of this run and the control in Figure \ref{fig:initialecc_ejectamass} to demonstrate this.  Both runs show the same double peak structure around 70 $\mjup$ (close to the hydrogen burning limit at 0.08 $\msol$), which reflects the original mass distribution of fragments (cf Figure \ref{fig:mhist_compare}).

\subsection{The Effect of Rapid Migration/Rapid Damping}

\begin{figure}
\begin{center}
\includegraphics[scale=0.4]{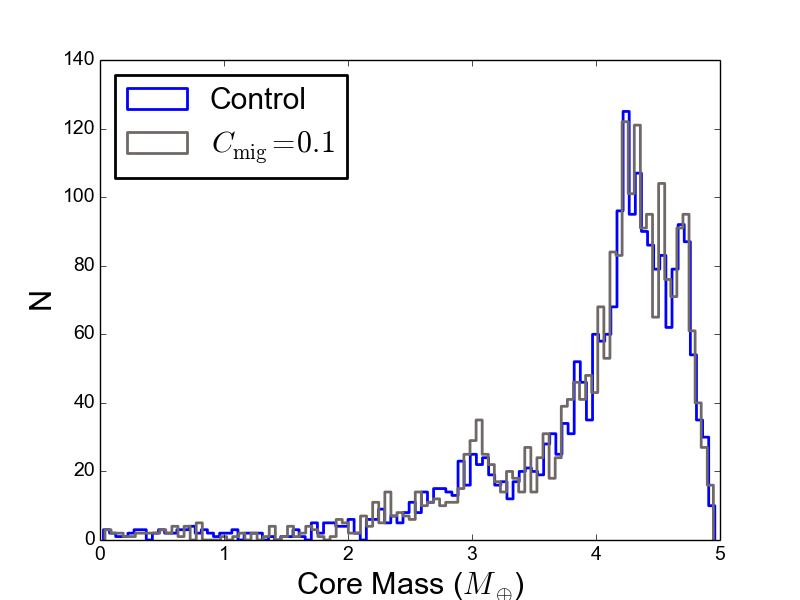}
\end{center}
\caption{The distribution of core masses for surviving bodies, in the control run, and for fast migrating runs with $\Cmig=0.1$ (runs C-on and R respectively). \label{fig:cmig_mcore}}
\end{figure}

\begin{figure}
\begin{center}
\includegraphics[scale=0.4]{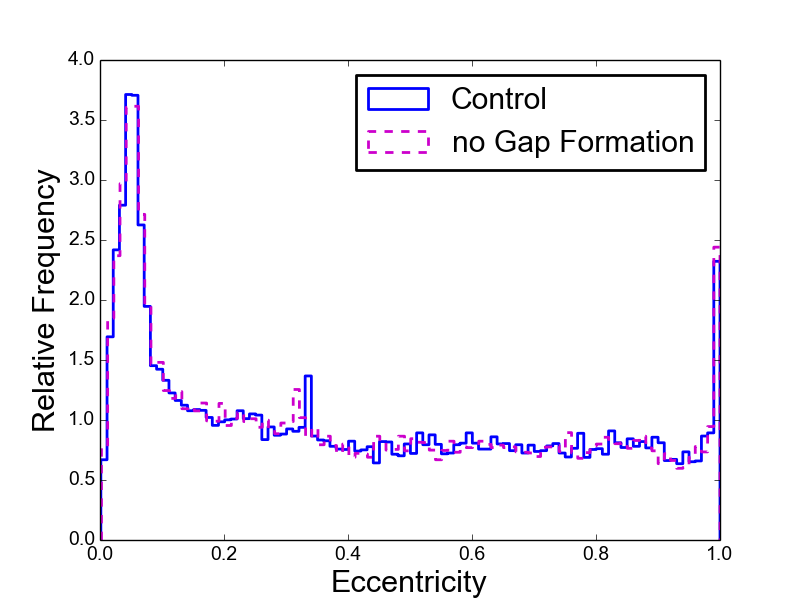}
\end{center}
\caption{The distribution of eccentricities for bound bodies, in the control run and for runs where gaps are not permitted to form (runs C-off and G respectively). \label{fig:nogaps_ehist}}
\end{figure}

\begin{figure}
\begin{center}
\includegraphics[scale=0.4]{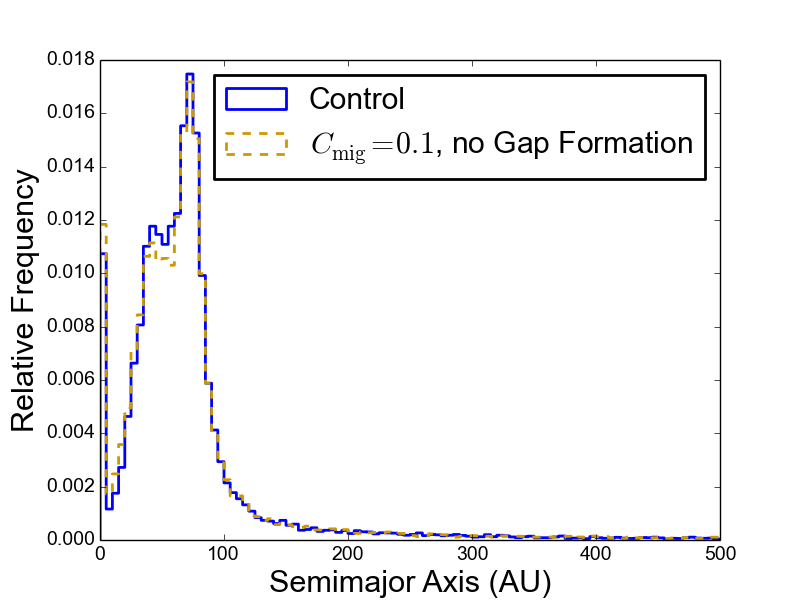}
\end{center}
\caption{The semimajor axis for bound bodies, in the control run and for runs where gaps are not permitted to form and $\Cmig=0.1$ (runs C-off and RG respectively). \label{fig:nogaps_ahist}}
\end{figure}

\noindent Studies of (single) migrating embryos in self-gravitating discs \citep{Baruteau2011,Malik2015a} have demonstrated that not only does Type I migration proceed at rates above predictions derived for non-self-gravitating discs, but the very act of establishing a gap and transitioning to the slower Type II regime is also frustrated\footnote{Equally, \citet{Stamatellos2015} demonstrated gap opening by accreting embryos in self-gravitating discs.  Whether the discrepancy is due to additional physics or differences in numerical formalism is not yet clear.}.  In the absence of fragment-fragment interactions, this would suggest that fragment destruction rates would be larger than those produced in population synthesis models relying on standard linear expressions for migration.  By the same token, it is possible - perhaps even likely - that self-gravitating discs would exert greater eccentricity and inclination damping forces.

To this end, we carried out a series of four model runs exploring the effects of i) reducing the migration timescale by setting $\Cmig = 0.1$, ii) preventing gap opening, iii) setting both $\Cmig=0.1$ and preventing gap opening and iv) reducing the eccentricity/inclination damping timescales (i.e. by reducing $\chi$ in equations \ref{eq:eccdamp} and \ref{eq:incdamp} from 10 to 1).  Each model run produced over 100,000 fragments, comprising around 30,000 unique planetary systems for each case.

In all four cases, we found the populations showed no significant differences from the control run, with the exception of a marginal increase in the fragment destruction rate.  For example, Figure \ref{fig:cmig_mcore} shows the resulting distribution of core masses in the control run, and the run where the migration timescale is reduced by a factor of ten.  Both runs produce a peak at approximately 4 $\mearth$ (with a lesser local maximum at 3$\mearth$).  These peaks are directly related to the double peak in the total mass distribution (Figure \ref{fig:mhist_compare}). The eccentricity distribution is governed by fragment-fragment scattering, not migration or damping (Figure \ref{fig:nogaps_ehist}).  

When we reduce both the migration timescale and suppress gap formation, the resulting semimajor axis distribution remains difficult to distinguish from the control (Figure \ref{fig:nogaps_ahist}).  The only exception is a slight increase in the number of objects in the nearest semimajor axis bin (and a slight deficit at around 40 AU).

Comparing this to models of planet formation via core accretion with planet-planet interactions, it is immediately clear that what \citet{Alibert2013} describe as the ``intermediate population'' at 50-100 au, which is largely absent in their model, remains in ours.  This is regardless of the specifics of the migration timescale or damping.  Equally, their models do produce objects at thousands of au and beyond, by similar scattering mechanisms to ours.  This suggests that both core accretion and disc instability systems must contribute at varying levels to the free floating planet population (see Discussion).

Throughout, we find very few examples of bodies entering orbital resonances or conducting convergent migration.  Core accretion population synthesis models commonly find almost all of their bodies in resonance if the number of initial embryos is low \citet{Rein2012,Alibert2013}.  These resonant chains, formed initially by convergent migration, can be broken for example during disc dispersal \citep{Izidoro2017}.  

As our fragments lose mass during inward migration, the migration rate tends to increase.  This positive feedback mechanism results in inner fragments accelerating away from the outer fragments, which prevents capture into a resonance.  Even without this acceleration, the systems formed are typically too dynamically unstable to enter resonance before suffering scattering events, and in some cases ejection.

\section{Discussion}
\label{sec:discussion}

\subsection{Limitations of the Analysis, and Directions for Future Work}

\noindent This paper has focused on the macrophysics of gravitational instability and disc fragmentation, in particular the dynamical consequences of multi-fragment systems.  However, we freely admit there are many other factors influencing the final population that our model has yet to incorporate.

Our fragment destruction rate remains at approximately 40\%, which is double that retrieved from \citet{Hall2017}.  We ascribe this to dynamical interactions not currently in the population synthesis model, principally the interaction between fragments and spiral structure.  Fragments passing through a spiral arm can experience stochastic outward ``kicks'', reducing the inward migration timescale.  This points to a larger issue in modelling the migration of multiple fragments in self-gravitating discs. As the spiral structure is modified and mediated by the tidal torques of fragments, the resulting interactions will result in a highly stochastic migration profile for each fragment, with each fragment's migration history being coupled to each other in a way that is difficult to model semi-analytically.

Once our fragments form, we do not accrete gas or solids from the surrounding disc.  As such, the total gas and dust mass of our fragments are underestimated.  This clearly has implications for the ability of our fragments to drive gaps \citep{Baruteau2011,Malik2015a}.  We are subsequently limited in our ability to model more sophisticated core formation modes, such as core assisted gas collapse \citep{Nayakshin2014a}, which may boost the number of objects with cores.  On a statistical level, we are limited in our ability to produce reliable metallicity correlations for our population \citep[cf][]{Nayakshin2015}.  

A population synthesis model that can reliably produce metallicity correlations would also require a more sophisticated disc model.  Throughout this work we have relied on interpolating a pre-evolved disc model, and assumed the existence of gaps through linear prescriptions.  The disc is also assumed to be a perfect mix of dust and gas, with a fixed dust-to-gas ratio of 0.01.  In reality, we should expect that the differential drag forces experienced by dust grains of varying sizes should mean that the dust and gas components of the disc eventually assume quite different profiles.  This is especially true if spiral density waves are present, which affect specific grain sizes at specific disc locations \citep{Rice2004,Clarke2009,Dipierro2015,Booth2016}.  The role of disc chemistry in affecting both grain and gas physics is also deserving of further exploration (but see \citealt{Ilee2011,Evans2015,Ilee2017}).

Our simulated disc does not self-consistently respond to fragment torques, and hence does not produce self-consistent gaps.  Fragment-fragment interactions are likely to be affected by the geometry of matter inside their horseshoe regions and the gaps they drive.  In other words, we should expect the interaction of multiple disc gaps to affect the orbital evolution of fragments.  It is this physics that makes the reversal in migration direction of Jupiter and Saturn when their gaps overlap after entering 3:2 resonance (the so-called ``Grand Tack'')  possible in models of Solar system formation \citep{Pierens2011,Walsh2012}.  

Are Grand Tacks possible in self-gravitating discs? The perturbations to disc structure caused by a fragment may induce further fragmentation \citep{Meru2015} so there is reason to believe that migration reversal might indeed be possible, although it remains unexplored.

What is more, the gaps that we prescribe preserve the fixed dust-to-gas ratio, and it is abundantly clear from simulations that conditions for gap formation in the dust do not in general match the conditions for gap formation in the gas \citep{Paardekooper2004,Dipierro2017}.

It is worth noting again that the high mass fragments produced in these models should result in strong non-axisymmetric perturbations on disc structure, and even shifts in the system centre of mass.  Our axisymmetric disc model does not respond to these perturbations, although the $N$-body elements in the system - the star and the fragments - do (thanks to our heliocentric formalism for the $N$-body algorithm).  It is likely that this mismatch will have dynamical consequences for all bodies in the system, and future work must attempt to incorporate the resulting $m=1$ modes induced in the disc by the star's motion around the system centre of mass.

These factors make it clear that future population synthesis models should incorporate a fully self-consistent disc model, ideally with two-fluid modelling of the dust and gas components.  Population synthesis with models such as this can be found in \citet{Nayakshin2015}, but not with multiple fragments present.    Their discovery of a dearth of giant planets \citep{Nayakshin2016a} may be partially or wholly explained by the lack of dynamical interactions.

We have fixed our outer disc radius at 100 au throughout this analysis - disc structures around Class 0/I objects exhibit a range of outer radii, some well beyond 100 au \citep[e.g.][]{Tobin2015,Johnston2015,Ilee2016}.  This clearly allows a larger number of possible fragments to form in individual systems, but it also allows fragments scattered from the inner system to still feel disc torques at larger radii, which would reduce the overall ejection rate.  Being able to run this model with a wider range of discs is highly desirable for future work.

A key finding of \citet{Hall2017} is the frequency of fragment mergers.  Merging fragments alter the mass function, and reduce the number of bodies in orbit without demanding an ejection.  Our N-Body modelling does not incorporate the tidal forces experienced by both bodies on close approaches, and it does not record impacts or grazing collisions, and subsequently does not capture the physics of fragment mergers.  We could imagine a variety of outcomes from such close approaches, from simple orbit modification (as we track in this work), to tidal dissipation in fragments during the encounter, resulting in atmospheric loss, to binary formation (with the potential formation of a circumbinary disc) and finally merging.  

We do not model individual fragment angular momenta, which is an impediment to studying these phenomena further.  Future work should focus on conducting high resolution hydrodynamic simulations of interacting fragments, to determine the possible rate of tidal stripping from fragment-fragment interactions, and to fully characterise the angular momentum evolution of both bodies.  This will also help us to understand the evolution of circumfragmentary discs, and whether they can form bound objects in orbit around the fragment.  While direct fragmentation of a circumfragmentary disc seems unlikely \citep{Forgan2016b}, the effects of tidal perturbations on such discs is unclear.

Indeed, our work does not compute any form of tidal force on the fragments, either between fragment pairs or between the fragment and the central star.  Inward scattering of fragments tends to produce a population of low semimajor axis, high eccentricity objects, just as is observed in population synthesis of core accretion systems.  We should expect tidal forces on these objects to be rather strong, reducing both semimajor axis and eccentricity.  

The secular timescales for these forces to act is much longer than our simulation runtime of 1 Myr.  Previous work has already shown that even an extra Myr of dynamical evolution has important consequences \citep{Li2015, TD_dynamics}, and we should take care when comparing the models in this paper to observations.  The subsequent migration behaviour of giant planets will change as the disc mass decreases and we enter the planet-dominated regime.  This points to further work focusing on longer simulation runtimes.

It is likely that many of the objects we form near the simulation's inner boundary are not fated to survive on Gyr timescales thanks to tidal dissipation.  Equally, the scattering of massive bodies followed by tidal circularisation can produce a population of bodies on close-in orbits, contaminating the inner exoplanet population.  This mechanism's overly high efficiency in producing hot Jupiters already points to disc fragmentation rarely forming planetary mass bodies \citep{Rice2015}.


\subsection{Implications for Observations of Bound Objects}

\noindent The most striking result of this work is the dominant role of scattering in sculpting the population of objects formed by GI.  This is almost counterintuitive when we consider that adding fragment-fragment interactions significantly increases the number of single-body systems.  Our attempts to strengthen the effects of migration (reducing the migration timescale, suppressing gap formation and guaranteeing the Type I migration regime at all masses) have failed to make any significant change to the population statistics.

This result relies on the initial multiplicity of fragmenting systems.  We have assumed that any fragmenting disc will form as many objects as can initially fit within the available space, which is defined by the Hill radius of the objects (especially the Hill radius of objects formed at the radial fragmentation boundary).  It is unlikely that every fragmenting disc will follow this rubric.  If a single fragment is formed, the evolution of said fragment will be governed by migration, not scattering. It is worth noting that we have considered a single maximum disc radius of 100 au, which effectively fixes the maximum initial multiplicity, and limits the effect of disc torques on scattered objects. 

It is also worth noting that \citet{Hall2017}'s set of simulations typically produce more than one fragment, and that a single fragment may trigger the formation of others \citep{Vorobyov2013,Meru2015}.  It is also worth noting that the initial separation of fragments is typically low (certainly no more than a few mutual Hill radii).  We would argue this guarantees scattering's importance in fragment evolution.

Fragment-fragment interactions tend to result in single object systems (with a few systems with multiplicity of 2, and the rest typically being empty systems).  Our data would suggest that future surveys attempting to find GI objects are likely to find single objects with masses near the planet/brown dwarf boundary.  We would interpret systems with a massive planet/brown dwarf at high semimajor axis, and a number of less massive bodies at low semimajor axis as evidence of GI and core accretion acting in the same system (\citealt{Boley2009}, but see also \citealt{Santos2017}).  An important data point in our understanding of how GI and core accretion cooperate in protostellar systems is the number of stars with brown dwarf companions and planetary mass bodies.  The most cited candidate for a system formed via GI, HR8799 \citep{Marois2008,Baines2012}, has a much higher multiplicity than we find here.  The stability of this system appears to have been assisted by resonant migration, requiring their formation at larger distances from the star \citep{Godziewski2014}.  Future work should explore how evolving more extended discs (around more massive stars) affects the multiplicity.

We produce a large number of bodies above a Jupiter mass with semimajor axes of a few hundred au - these would be easily resolved by direct imaging studies if the body is warm enough.  Detecting this population will provide important circumstantial evidence for/against disc fragmentation.  

\citet{Tobin2016}'s detection of companions embedded in the disc of a massive star is convincing evidence for fragmentation.  Our model would also indicate that Elias 2-27's observed spiral structure is consistent with disc fragmentation \citep{Meru2017}, but would also suggest that a fragment will require dynamical kicks from the spiral structure or another companion to survive for long times.


\subsection{Implications for Observations of Free Floating Objects}

\noindent Our data shows the free floating planet population produced by GI will have a mass function that is double-peaked, with a relatively low mean velocity of around $5 \kms$ (and a dispersion of around $2 \kms$).  The ejecta velocity distribution relatively insensitive to the initial separation of fragments, but higher fragment spacing tends to produce more massive fragments (both bound and ejected), as the Jeans mass increases with distance from the star.  

The observed mass function and velocity distributions will be a blend of the core accretion statistics \citep[e.g.][]{Veras2012} and the GI statistics we present here.  A measurement of these distributions down to masses less than $1 \mjup$ will give the first indications of the relative frequency of planet formation via GI versus planet formation via core accretion.

Our ejection rates are actually an underestimate - the large population of bodies with semimajor axes beyond 500 au are very likely to be stripped from the system in typical young cluster environments, although our previous work in this area suggests the resulting change to the ejecta mass function will be minimal \citep{TD_dynamics}.

However, the same work shows that the effect of adding a cluster potential will also drive strong changes in the orbital inclination.  Even if objects remain bound (or are ejected and are re-captured), we should expect to observe larger variation in orbital alignments than seen in our data.

Given that these ejections occur at early times, we should also expect  ejected objects to retain circumplanetary discs.  The recent detection of a disk around the 12 $\mjup$ object OTS44 \citep{Bayo2017} is an interesting candidate for an ejected disc fragment.  The disc-to-object mass ratio tends to obey scaling relations for low mass stars, but this relies on a dust mass estimated from continuum flux, which assumes the disc to be optically thin.  Such an assumption can be hazardous if the disc is optically thick \citep[cf][]{Forgan2013a,Forgan2016e}, which we might expect to be the case for circumfragmentary discs.  Objects such as OTS44 are important testbeds for the evolution of fragment angular momentum both before and after ejection from its natal disc.

Equally, OTS44 may have been formed via core accretion, and then ejected.  Simulations of the formation of circumplanetary discs indicate that subdiscs around objects formed via GI are significantly cooler than their counterparts formed via core accretion \citep{Forgan2016b,Szulagyi2017}.  Circumfragmentary discs possess significantly higher total angular momentum budgets, and are largely Toomre stable with relatively long lifetimes.  If the disc survives the ejection process, determining the disc temperature profile would be a key piece of evidence for OTS44's formation history.


\section{Conclusions}
\label{sec:conclusions}

\noindent We have continued our development of fully consistent population synthesis models of disc fragmentation via gravitational instability (GI), and the subsequent tidal downsizing of disc fragments.

Our latest model now includes the effects of fragment-fragment interactions while the disc is still present.  We see that such interactions produce a profound change in the population of gas giant planets and brown dwarfs that we produce.  The scattering effect of fragment-fragment interactions sculpts the population, dominating over migration effects (even when the effects of migration are enhanced).

Scattering reduces the multiplicity of systems formed via GI, and increases the number of systems with no surviving objects.  Typical systems formed by GI consist of one or two gas giant planets or brown dwarfs, with relatively low eccentricity, and semimajor axes above 30 AU.  

In contrast to our previous work, we find a small increase in the number of low mass rocky bodies formed, including disrupted fragments producing planetesimal belts.  We stress however that these outcomes remain extremely rare, and that the survival of terrestrial planets formed via GI is uncertain due to their low semimajor axis and high eccentricity.

We still find that around 40\% of all fragments are destroyed, which is still large compared to hydrodynamic simulations (which we ascribe to the effect of nonaxisymmetric disc structure).  Of the surviving bodies, a further 38\% (i.e. around 20\% of all fragments formed) are ejected, with relatively low initial velocities around 5 $\kms$.  This ejection rate does not consider external perturbations from Galactic tides or nearby clusters, and so is an underestimate.

Importantly, we find that the population synthesis model continues to produce predominantly massive objects at large semimajor axis, which should be observed by direct imaging surveys, with an extra component at lower semimajor axis, lurking in the parameter space typically occupied by planets formed via core accretion.  The expected number of planets formed via GI at low semimajor axis remains unclear, as simulations with longer runtimes are needed to determine their secular evolution on Gyr timescales.  We should also note that population synthesis of GI is still in its infancy.  The evolution of turbulent dust-gas mixtures during the fragmentation process requires further study, and other dynamical effects due to disc asymmetry and spiral structure are yet to be incorporated in our models.

The few detections to hand from direct surveys are consistent with GI \citep{Vigan2017}.  Our current models suggest that objects formed via GI are likely to remain a small contaminant in the exoplanet population, with the majority of observed exoplanets being formed by core accretion.

\section*{Acknowledgements}

DHF gratefully acknowledges support from the ECOGAL project, grant agreement 291227, funded by the European Research Council (ERC) under ERC-2011-ADG.  This project has received funding from ERC under the European Union's Horizon 2020 research and innovation programme (grant agreement No 681601).  The research leading to these results also received funding from the European Union Seventh Framework Programme (FP7/2007-2013) under grant agreement number 313014 (ETAEARTH).  KR acknowledges the support of the UK Science and Technology Facilities Council through grant number ST/M001229/1.   FM acknowledges support from The Leverhulme Trust and the Isaac Newton Trust.  The authors warmly thank the anonymous reviewer for their insightful reading of this manuscript.  This research  has  made  use  of  NASA's  Astrophysics  Data  System Bibliographic  Services.  This work relied on the compute resources of the St Andrews MHD Cluster.




\bibliographystyle{mnras} 
\bibliography{TD_nbody}


\bsp	
\label{lastpage}
\end{document}